\documentclass[twocolumn, titlepage, superscriptaddress, amsmath, amssymb, 10pt, floatfix]{revtex4-2}
\usepackage[english]{babel}
\usepackage[utf8]{inputenc}
\usepackage[colorinlistoftodos, color=green!40, prependcaption]{todonotes}
\usepackage{amsthm}
\usepackage{mathtools}
\usepackage{physics}
\usepackage{xcolor}
\usepackage{graphicx}
\usepackage[left=23mm,right=13mm,top=35mm,columnsep=15pt]{geometry} 
\usepackage{adjustbox}
\usepackage{placeins}
\usepackage[T1]{fontenc}
\usepackage{lipsum}
\usepackage{csquotes}
\usepackage{verbatim}
\usepackage{tikz}
\usepackage{qcircuit}

\newcommand{\be}{\begin{equation}}
\newcommand{\ee}{\end{equation}}

\newcommand{\psimps}{\psi_{\chi}}
\usepackage[pdftex, pdftitle={Article}, pdfauthor={Author}]{hyperref} 
\begin{document}
\title{Pre-optimizing variational quantum eigensolvers with tensor networks}

\author{Abid Khan}
\affiliation{Department of Physics, University of Illinois Urbana-Champaign, Urbana, IL, United States 61801}
\affiliation{USRA Research Institute for Advanced Computer Science (RIACS), Mountain View, CA, 94043, USA}
\affiliation{NASA Ames Research Center, Moffett Field, CA, 94035, USA}
 \author{Bryan K. Clark}
\affiliation{Department of Physics, University of Illinois Urbana-Champaign, Urbana, IL, United States 61801}
\affiliation{IQUIST and Institute for Condensed Matter Theory and NCSA Center for Artificial Intelligence Innovation, University of Illinois at Urbana-Champaign, IL 61801, USA}
\author{ Norm~M.~Tubman}
\affiliation{NASA Ames Research Center, Moffett Field, CA, 94035, USA}

\date{\today} 

\begin{abstract}
The variational quantum eigensolver (VQE) is a promising algorithm for demonstrating quantum advantage in the noisy intermediate-scale quantum (NISQ) era.  However, optimizing VQE from random initial starting parameters is challenging due to a variety of issues including barren plateaus, optimization in the presence of noise, and slow convergence.   While simulating quantum circuits classically is generically difficult, classical computing methods have been developed extensively, and powerful tools now exist to approximately simulate quantum circuits.  This opens up various strategies that limit the amount of optimization that needs to be performed on quantum hardware. Here we present and benchmark an approach where we find good starting parameters for parameterized quantum circuits by classically simulating VQE by approximating the parameterized quantum circuit (PQC) as a matrix product state (MPS) with a limited bond dimension. 
Calling this approach the variational tensor network eigensolver (VTNE), we apply it to the 1D and 2D Fermi-Hubbard model with system sizes that use up to 32 qubits. 
We find that in 1D, VTNE can find parameters for PQC whose energy error is within 0.5\% relative to the ground state.  In 2D, the parameters that VTNE finds have significantly lower energy than their starting configurations, and we show that starting VQE from these parameters requires non-trivially fewer operations to come down to a given energy. The higher the bond dimension we use in VTNE, the less work needs to be done in VQE. By generating classically optimized parameters as the initialization for the quantum circuit one can alleviate many of the challenges that plague VQE on quantum computers.

\end{abstract}



\maketitle

\section{Introduction} \label{sec:introduction}

The variational quantum eigensolver (VQE) is particularly well-suited for the noisy intermediate-scale quantum (NISQ) regime, where quantum computers are limited in size and coherence time. Some advantages of VQE are that its variational character can provide some degree of error mitigation in the parameterization of the gates~\cite{m1, m2, m3, m4} and that it features shallower circuits compared to more exact algorithms such as phase estimation and quantum approximate optimization algorithms (QAOA)~\cite{s1, s2, s3, s4, s5, a12020034, PhysRevA.97.022304, sureshbabu2023parameter, PhysRevA.107.062406, farhi2014quantum, kremenetski2023quantum, kremenetski2021quantum,tubman2018postponing}. The applications for VQE range over a number of different fields~\cite{sawaya2023hamlib} including chemistry~\cite{s4, s2, m3, c1, c2, c3, c4, mullinax2023largescale}, materials science~\cite{s1, ms1}, and machine learning~\cite{ml1, ml2, ml3}. 

Despite the promise of parameterized quantum algorithms to provide advantages over classical methods~\cite{qa1, PhysRevApplied.19.024047}, several obstacles obstruct their realization. In particular, the parameterized quantum circuit (PQC) optimization landscape is plagued by the presence of barren plateaus~\cite{bp1, bp2, bp3, bp4, bp5}, 
particularly starting from randomly parameterized quantum circuits,
and local minima~\cite{lm1, lm2, lm3}. These problems have been explored in quantum chemistry applications, where circuits for computing molecular ground states can reach high-precision results using initializations based on mean-field Hartree-Fock or more sophisticated coupled-cluster-based solutions~\cite{s2,qc1,tubman2018postponing,hirsbrunner2023mp2}. There is active work on mitigating these challenges and finding ways to improve the performance of VQE~\cite{bp2, bp3}.

Another difficulty in demonstrating an advantage over classical algorithms using PQCs is the increasing sophistication of classical simulation algorithms.   However, this also provides new possibilities for performing pre-optimization on classical hardware. 
 Several ideas of this type have been suggested recently for applications in quantum chemistry~\cite{baek2022say} and in related works these ideas were considered in detail with large-scale simulations up to 64 qubits~\cite{mullinax2023largescale,hirsbrunner2023mp2}.  To perform simulations at these scales, interesting approximations for simulation of quantum circuits have to be employed~\cite{free2022,mullinax2023largescale,hirsbrunner2023mp2}.  While the above simulations  have been performed at large scales, there are many alternative approaches in which these ideas can be further explored which include tensor network approaches~\cite{tn1, tn2}. The ability of tensor networks to be deployed on powerful classical hardware accelerators, such as graphical and tensor processing units (GPUs and TPUs), raises the bar for quantum hardware to overcome. 

Rather than perceiving this as an obstacle to quantum advantage, one can instead view the success of sophisticated classical simulation techniques, including tensor network algorithms, as a path towards realizing quantum advantage: we can first classically simulate PQC optimization and then continue the work on quantum hardware. In this paper, we call this approach of classical optimization the variational tensor network eigensolver (VTNE).
With VTNE, we bridge the gap between classical and quantum optimization by demonstrating how to find a good set of intermediate parameters for the VQE circuit by first approximately classically optimizing the VQE using tensor networks.

\section{Methods} \label{sec:methods}
We use MPSs as the tool for efficient quantum circuit simulation and optimization.
To accomplish our goal of using MPSs as a pre-optimization tool, we start with a fixed gate structure on the quantum hardware, as might naturally be dictated by the physical device.  
For each set of parameters of the quantum circuit, we map it onto an approximate MPS.
This MPS realization of the quantum circuit is then the starting point from which we compute an approximate energy and its derivative given the circuit parameters. 
The MPS is generated by starting with a tensor network where 
each unitary gate within the circuits is translated into a rank $2d$ tensor, where $d$ represents the number of qubits the gate interfaces with, which is then contracted into an MPS.


\begin{figure}[!htbp]
    \centering
    \includegraphics[width=\linewidth]{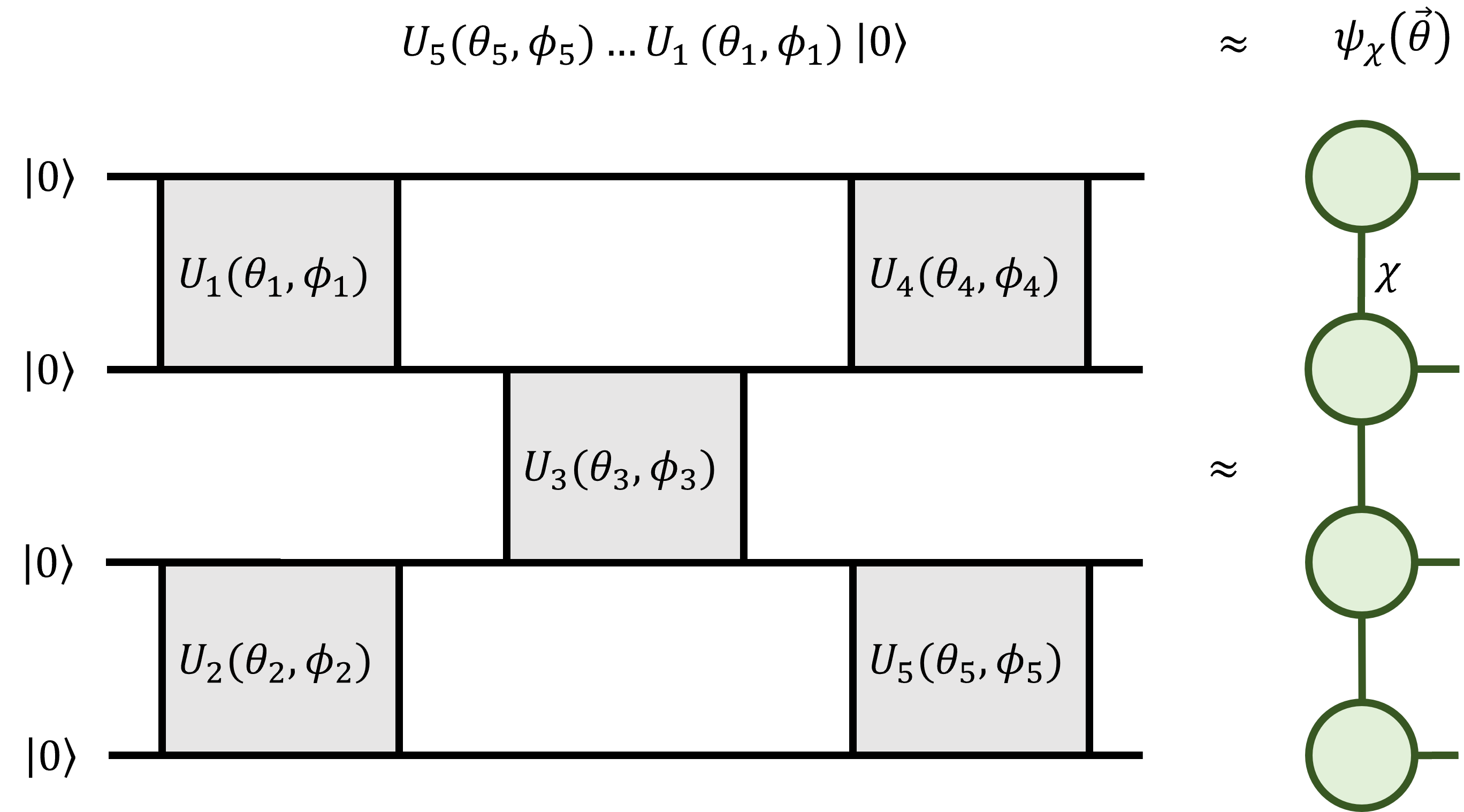}
    \caption{Contraction of a quantum circuit $\ket{\psi_{\text{PQC}}}$ into an MPS $\ket{\psimps}$. With the contraction of each gate, the bond dimension is capped at $\chi$.}
    \label{fig:circ_to_mps}
\end{figure}

The task of transforming a state obtained from a PQC into an MPS form presents challenges for highly entangled states due to the bond dimension of the MPS. To address this, we approximate the quantum circuit as an MPS with a fixed bond dimension $\chi$ significantly smaller than the maximum bond dimension $\chi_{\text{max}} = 2^{n_{q}/2}$, where $n_{q}$ is the number of qubits. 
We then examine how the improved classical starting configuration depends on the bond dimension of the MPS which controls the computational complexity of the classical optimization. 
Starting with a good point, this approach may help alleviate the difficulties of executing VQE on quantum computers and set the stage for a more explicit demonstration of quantum advantage.
\subsection{Model}
In this work, we consider the one and two-dimensional Fermi-Hubbard model. This model is particularly interesting because its regular structure and relatively simple form suggest that it may be easier to implement on NISQ devices~\cite{arute2020observation}. We anticipate the high-level approaches we introduce will also apply to other condensed-matter systems.  The Hamiltonian for the Hubbard model is 
\begin{equation} \label{eq:hubbard} 
    H = -t \sum_{\langle i, j \rangle,\sigma\in \{\uparrow, \downarrow\}} (a_{i\sigma}^\dag a_{j\sigma} + a_{j\sigma}^\dag a_{i\sigma}) + U \sum_i n_{i\uparrow}n_{i\downarrow}, 
\end{equation}
where $a_{i\sigma}^\dag$, $a_{i\sigma}$ are fermionic creation and annihilation operators; $n_{i\sigma} = a_{i\sigma}^\dag a_{i\sigma}$ and $\langle i, j \rangle$ correspond to nearest neighbors, $t$ is the nearest neighbor hopping,  and $U$ is the on-site potential. Throughout the rest of this paper, we use $t = 1$ and $U=2$ and work in the half-filled regime. We use the well-known Jordan-Wigner encoding of the fermionic Hamiltonian as a qubit Hamiltonian~\cite{Barnes_2001}.

\subsection{Ansatz}
\begin{figure}[!htbp]
     \centering
    \includegraphics[width=\linewidth]{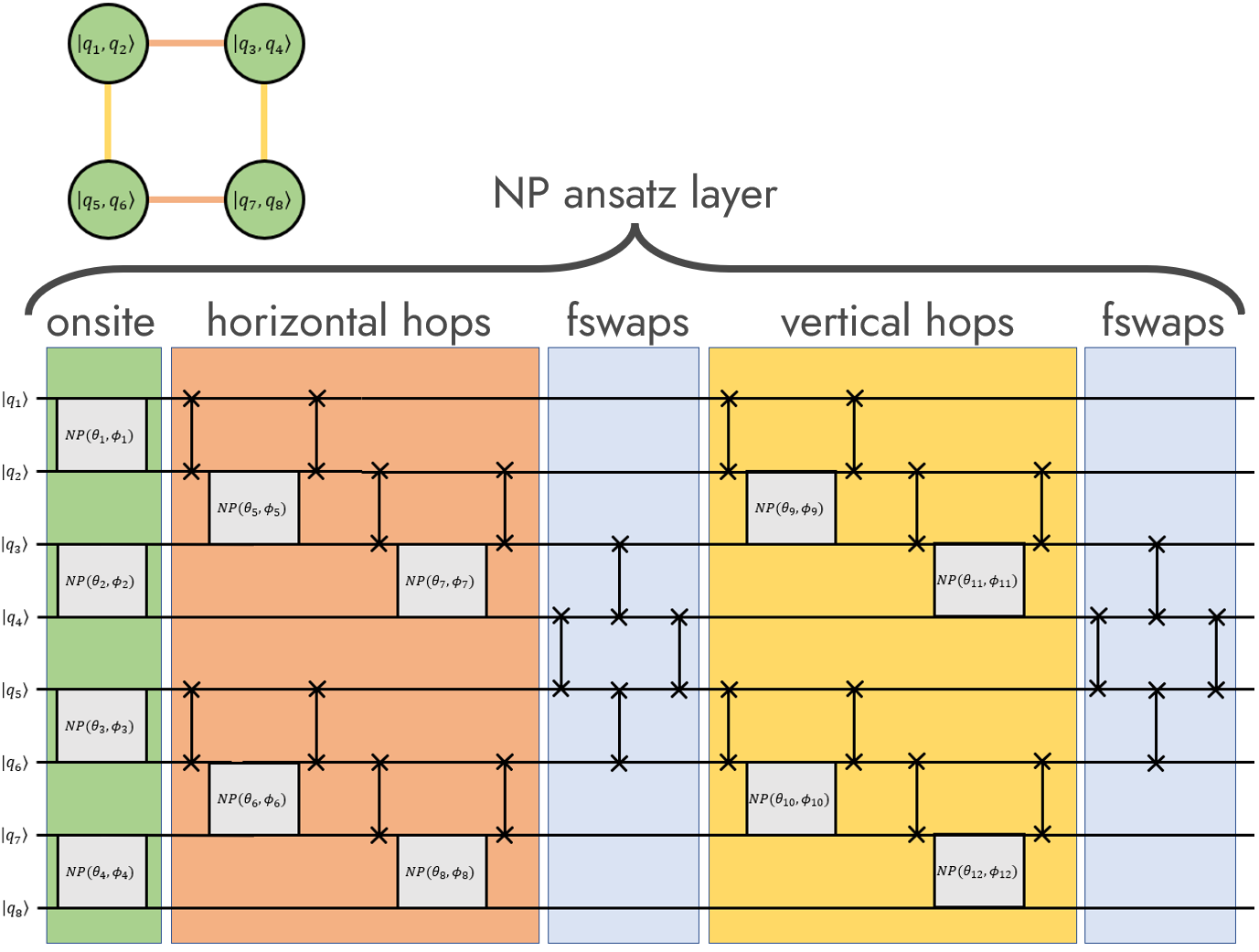}
    \caption{One layer of the number-preserving ansatz for the $2\times 2$ lattice. Consisting of number-preserving gates defined in Eq.~\ref{eq:np_gate} and fermionic swap gates defined in Eq.~\ref{eq:fswap}, the layer starts with onsite interacting gates, followed by horizontal hopping gates, where fermionic swap operators are used to bring columns-wise nearest-neighbor same-spin wires together. Then, fermionic swaps are performed to bring rows-wise nearest neighbor same-spin wires together. Finally, fermionic swaps are used to return all the wires to their original positions.}
    \label{fig:np_ansatz}
\end{figure}

We consider the number-preserving ansatz  used in ref.~\cite{s1}, which in the case of the Fermi-Hubbard model, is more general than its associated Hamiltonian variational ansatz (HVA)~\cite{qc1}. This ansatz consists of a parameterized number-preserving  gate
\begin{equation}
    NP(\theta, \phi) = \begin{pmatrix}
                            1 & 0 & 0 & 0 \\
                            0 &  \cos{\theta} & -i\sin{\theta} & 0 \\
                            0 & i\sin{\theta} &   \cos{\theta} & 0 \\
                            0 &             0 &              0 & e^{i\phi} \\
                        \end{pmatrix}
    \label{eq:np_gate}
\end{equation}
and a parameter-less fermionic swap gate
\begin{equation}
       \text{FSWAP}=   \begin{pmatrix}
                        1 & 0 & 0 &  0 \\
                        0 & 0 & 1 &  0 \\
                        0 & 1 & 0 &  0 \\
                        0 & 0 & 0 & -1 \\
        \end{pmatrix}.
    \label{eq:fswap}
\end{equation}
This circuit consists of qubits, labelled as $(i,j,\sigma)$, patterned on a 1D or 2D lattice where $(i,j)$ specifies a lattice site position and $\sigma\in \{\uparrow,\downarrow\}$.  $(i,j,\uparrow)$ is always directly to the right of $(i,j,\downarrow)$.
A layer of this ansatz starts with a set of two-qubit gates interacting between $(i,j,\uparrow)$ and $(i,j,\downarrow)$.
Following this, we have horizontal and vertical hopping gates between the four commuting sets of hopping terms $(i,j,\sigma)\leftrightarrow (i,j+1,\sigma)$; $(i,j,\sigma)\leftrightarrow (i+1,j,\sigma)$; $(i,j+1,\sigma)\leftrightarrow (i,j+2,\sigma)$; $(i+1,j,\sigma)\leftrightarrow (i+2,j,\sigma)$; where $i$ and $j$ are even.  Between each set, a group of fermionic swaps is performed so that the corresponding sites being hopped between are consecutive. 
Illustrated in figure~\ref{fig:np_ansatz} for a $2\times 2$ lattice, the number-preserving ansatz has shown success in capturing the ground state energy of the Fermi-Hubbard model for up to 24 qubits~\cite{s1}. We prepend this ansatz in our simulations with $R_{z}(\theta)$ gates at each qubit. In testing, we found this layer helped to improve optimization.

\subsection{MPS from Quantum Circuit}
Our goal now is to find good parameters for the variational circuit classically. To accomplish this, we will need to approximate the VQE circuit as an MPS, which is done as follows.  Let
\begin{equation}
    \ket{\psi_{\text{PQC}}\qty(\va{\theta})} = U_{n}(\vec{\theta}_{n})\ldots U_{2}(\vec{\theta}_{2})U_{1}(\vec{\theta}_{1})\ket{\psi_0}
\end{equation} 
where $\ket{\psi_0}$ is an unparameterized starting state.
We approximately represent this wave-function 
 as a bond dimension $\chi$ MPS $\ket{\psimps(\vec{\theta})}$ (See  figure~\ref{fig:circ_to_mps}.)  
 The approximation is performed via the time-evolving block decimation (TEBD) technique~\cite{PhysRevLett.93.040502}. 
 This construction provides us with a truncated representation of our PQC, so that 
 we can classically optimize the energy function
\begin{equation}\label{eq:Echi}
    E_{\chi}(\vec{\theta}) = \bra{\psimps(\vec{\theta})}H\ket{\psimps(\vec{\theta})},
\end{equation}
where $H$ is
represented as a matrix product operator (MPO). Generally, k-local Hamiltonians have a simple MPO implementation~\cite{ORUS2014117, Bridgeman_2017}. As we increase $\chi$, the optimized energy $E_\chi$ gets closer to the exact energy $E_\text{exact}$
\begin{equation}\label{eq:Eexact}
    E_{\text{exact}}(\vec{\theta}) = \bra{\psi_\text{PQC}(\vec{\theta})}H\ket{\psi_\text{PQC}(\vec{\theta})}.
\end{equation}
Note that $ E_{\text{exact}}(\vec{\theta}) = E_{\chi_{\text{max}}}(\vec{\theta})$, where $\chi_\text{max}=2^{n_{q}/2}$. 
Throughout this paper, we use the ITensor package~\cite{ITensor} to compute all our tensor network calculations.
\subsection{Optimization}
Given a bond dimension $\chi$, the objective function that we optimize is Eq.~\ref{eq:Echi}. We begin our optimization by finding the ground state of the non-interacting $(U=0)$ case. 
We optimize two non-interacting number-preserving ansatze; one in which spins occupy the even sites (for the spin-up determinant) and one where the spins occupy the odd sites (for the spin-down determinant) so that when we consider the full interacting system, the state starts with a checkerboard of up and down spin configurations.  In this optimization, we need only half the qubits and can remove both the onsite gates, any swap gates required to cross over different flavored spins, and any hopping terms on the different flavored spins.  This leaves less the half the number of parameters to optimize. 
 We initialize the parameters using a Gaussian distribution $\mathcal{N}\qty(0, 10^{-5})$, and we carry out the minimization using the Broyden-Fletcher-Goldfarb-Shannon (BFGS) algorithm~\cite{10.1093/imamat/6.1.76, 10.1093/comjnl/13.3.317}. We terminate the optimization when the function tolerance reaches $10^{-7}$ or the gradient norm reaches $10^{-6}$. Note that if $\chi < 2^{(n_{x}n_{y}/2)}$, this optimization will not necessarily give us parameters representing the exact ground state for the non-interacting case. After performing the non-interacting optimization, we use those parameters to start the optimization for the interacting ($U=2$) case. 

\begin{figure*}[!htbp]
    \centering
    \includegraphics[width=\linewidth]{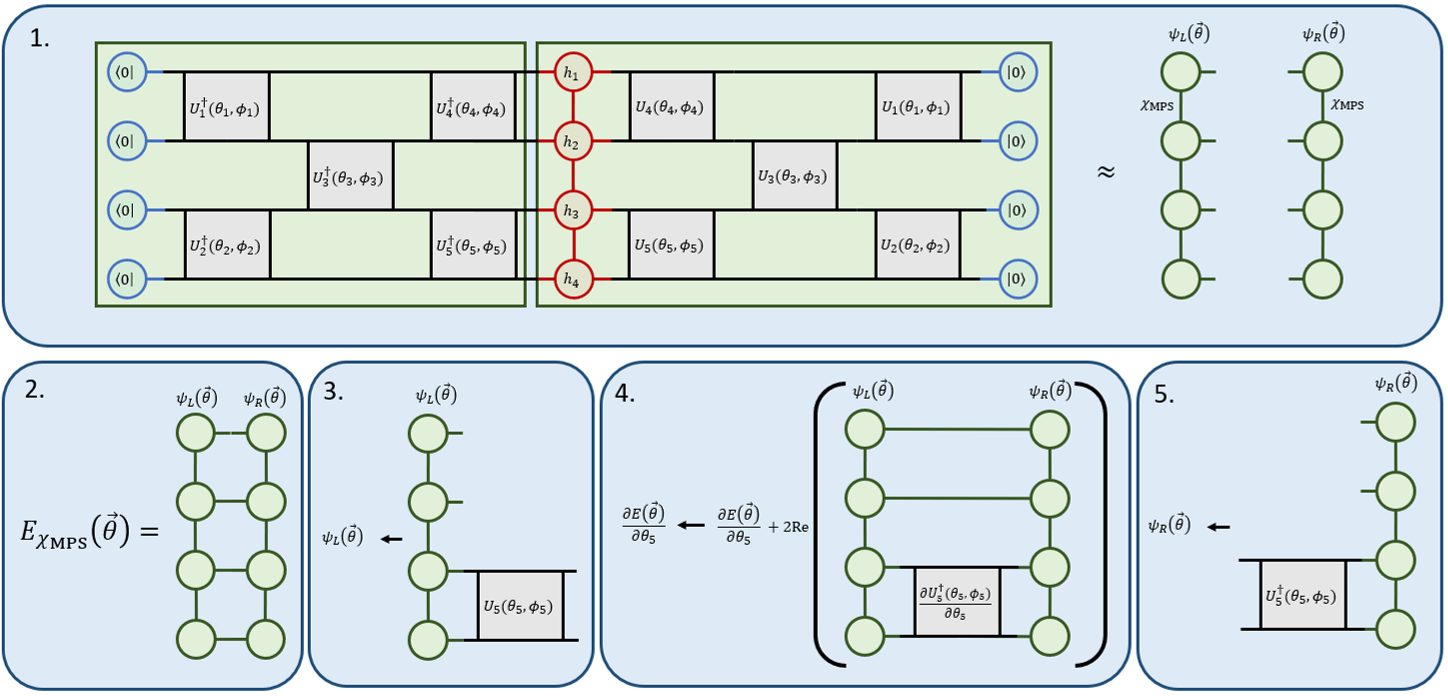}
    \caption{Procedure for computing the approximate energy (1-2) $E_{\chi_{\text{MPS}}}(\vec{\theta}) = \bra{\psimps (\vec{\theta})}H\ket{\psimps(\vec{\theta})}$ and (3-5) its gradient. (1) We first construct two MPSs $\psi_{L}\approx\bra{0}U_{1}^{\dagger}(\vec{\theta}_{1})U_{2}^{\dagger}(\vec{\theta}_{2}) \ldots U_{n}^{\dagger}(\vec{\theta}_{n})$ and $\psi_{R}\approx H U_{n}(\vec{\theta}_{n})\ldots U_{2}(\vec{\theta}_{2})U_{1}(\vec{\theta}_{1})\ket{0}$. (2) The approximate energy is then the contraction of these two MPSs $E_{\text{MPS}} = \braket{\psi_{L}}{\psi_{R}}$. (3) We pop out the last gate from $\psi_{L}$ by applying the inverse to it. (4) We then compute gradient terms with respect to the parameters in the gate by performing the contraction $\bra{\psi_{L}}\vec{\nabla} U_{n}(\vec{\theta}_{n})\ket{\psi_{R}}$. (5) We move the popped gate from $\psi_{L}$ to $\psi_{R}$. We repeat (3-5) until $\psi_{L}$ is parameter-less.}
    \label{fig:grad}
\end{figure*}

Classical techniques exist  to efficiently compute the gradient of Eq.~\ref{eq:Echi}, which include automatic differentiation (AD)~\cite{PhysRevX.9.031041}.  Here, we implement an approximate gradient scheme that does not require as much memory and only needs two circuit evaluations. 
To derive our approximation, we start with the gradient of the exact energy of the full PQC. 
For a unitary $U_{k}(\vec{\theta}_{k})$ containing a parameter $\theta_{k,i}\in \vec{\theta}_{k}$, the derivative of the energy with respect to that parameter is
\begin{equation}\label{eq:gradE}
 \frac{\partial E_\text{exact}(\vec{\theta})}{\partial \theta_{k,i}} = 2\text{Re}\left[  \mel**{\psi^{(k)}_{L}}{\frac{\partial U_{k}^{\dagger}(\vec{\theta}_{k})}{\partial \theta_{k, i}}}{\psi^{(k)}_{R}}\right],
\end{equation}
where
\begin{align}
\bra{\psi^{(k)}_{L}} &= \bra{0}U_{1}^{\dagger}(\vec{\theta}_{1})\ldots U_{k-1}^{\dagger}(\vec{\theta}_{k-1})\\
\ket{\psi^{(k)}_{R}} &= U_{k+1}^{\dagger}(\vec{\theta}_{k+1}) \ldots U_{n}^{\dagger}(\vec{\theta}_{n}) H \ket{\psi_{\text{PQC}}}
\end{align}
We can then iteratively compute the derivative with respect to each parameter by updating $\bra{\psi^{(k)}_{L}}$ and $\ket{\psi^{(k)}_{R}}$ recursively:
\begin{align}
\bra{\psi^{(k-1)}_{L}} &= \bra{\psi^{(k)}_{L}} U_{k-1}(\vec{\theta}_{k-1})\\
\ket{\psi^{(k-1)}_{R}} &= U_{k}^{\dagger}(\vec{\theta}_{k})\ket{\psi^{(k)}_{R}}
\label{eq:psilr}
\end{align}
Figure~\ref{fig:grad} depicts this process through tensor network diagrams. This gradient computation is exact in the limit of the maximal bond dimension. In our MPS approximation regime, we truncate $\bra{\psi^{(k-1)}_{L}}$ and $\ket{\psi^{(k-1)}_{R}}$ in Eq.~\ref{eq:psilr} to bond dimension $\chi$ after each iteration.  This derivative approximates the derivative of Eq.~\ref{eq:Eexact}, which is not the same as the derivative of Eq.~\ref{eq:Echi}.

\section{Results} \label{sec:results}

We start by looking at the results for the 1D and 2D Hubbard models at various lattice sizes. For each lattice size $(n_{x}, n_{y})$, the number of layers used per lattice configuration was chosen by using the results of ~\cite{s1} which provided depths that led exact-VQE  to 0.99 fidelity with the true ground state for systems up to 24 qubits. We used those circuit depths, and for systems with more than 24 qubits, we linearly extrapolated from the depth vs qubit data. Table~\ref{tab:configs} shows the lattice configurations used along with the number of layers used in the number-preserving ansatz. 

\begin{table}[!htbp]
\centering
\begin{tabular}{|c|c|c|c|}
\hline
$n_{x}$ & $n_{y}$ & qubits & layers \\
\hline
 4 & 1 &  8 &  4 \\
 8 & 1 & 16 &  7 \\
12 & 1 & 24 & 11 \\
16 & 1 & 32 & 14 \\
 4 & 2 &  8 & 10 \\
 4 & 3 & 24 & 17 \\
 4 & 4 & 32 & 24 \\
\hline
\end{tabular}
\caption{A table of lattice configurations with the number of layers used for the number-preserving ansatz.}
\label{tab:configs}
\end{table}

\begin{figure}[!htbp]
    \centering
    \includegraphics[width=\linewidth]{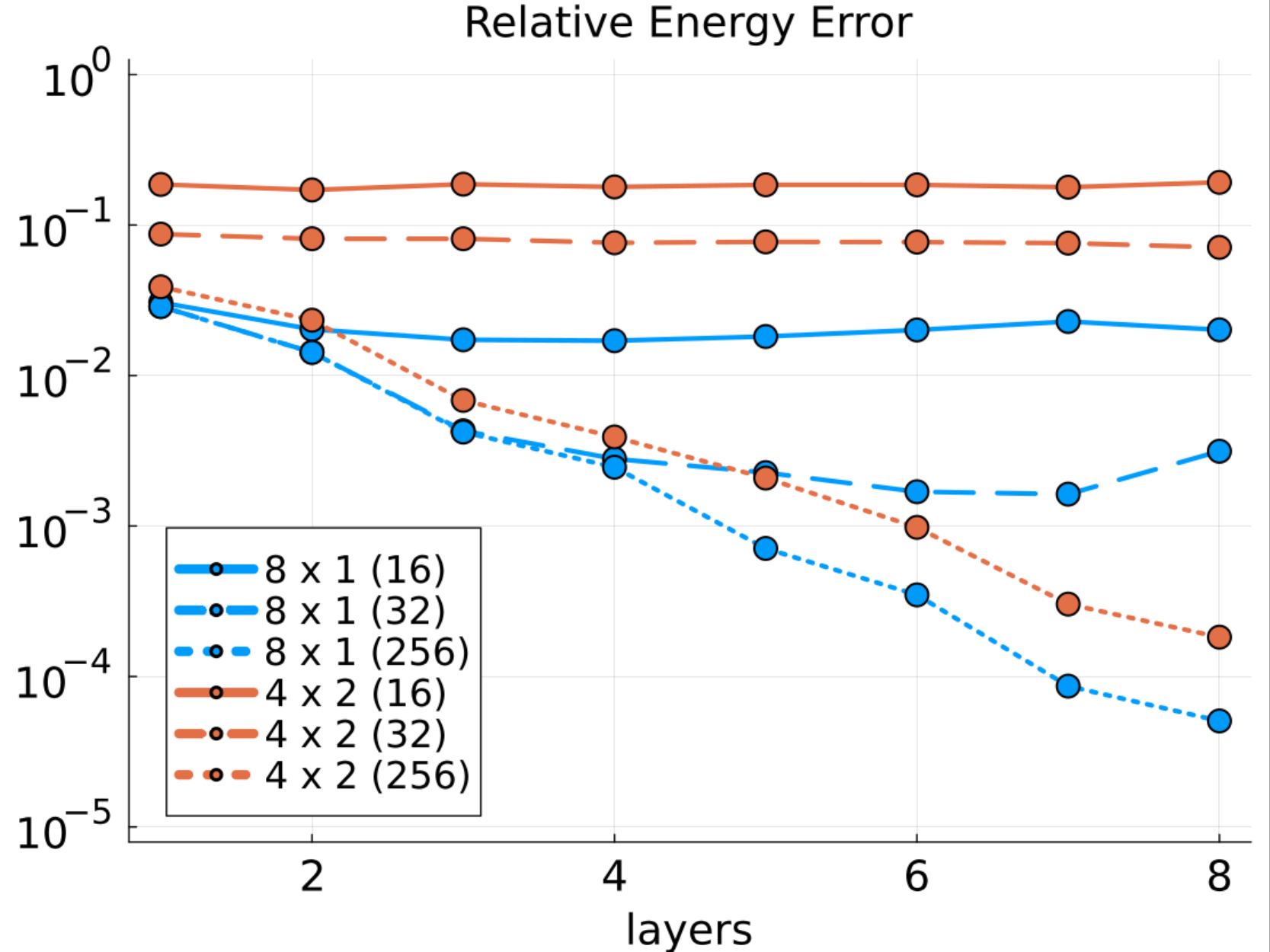}
    \caption{Relative energy error vs layer depth for a $8\times1$ (blue) and $4\times2$ (red) Hubbard model for bond dimensions $\chi=16$ (solid lines), $\chi=32$ (dashed lines), and the full bond dimension $\chi_\text{exact}=256$ (dotted lines).}
    \label{fig:depth}
\end{figure}

Figure~\ref{fig:depth} shows the relative energy error of the ground state as we increase the layer depth for $1\times 8$ and $2\times 4$ system sizes at different bond dimensions. When the VQE is performed using the full bond dimension $(\chi = 256)$, increasing the number of layers leads to more accurate ground state energies. Figure~\ref{fig:depth} also tells us for a given depth, how well we optimize at a certain bond dimension. In the case of the 1D (8x1) system, we find that at one layer, optimizing at bond dimension 16 is enough to completely optimize our ansatz. This is evident by observing that optimizing at bond dimensions 32 and 256 yields the same relative energy error as with bond dimension 16. When we add another layer, a bond dimension of 16 is no longer enough to represent the state exactly, but optimizing at a bond dimension of 32 is. Once we are at 4 layers the bond dimension 32 optimization is no longer able to exactly represent the PQC. Overall, we see that as we add layers to the ansatz, we require more bond dimension to more accurately represent our PQC and drive our optimization down. Turning to the 2D (4x2) system, we find that neither bond dimensions 16 nor 32 are enough to fully represent the PQC even for one layer.  

VQE optimizations are performed by capping the MPS bond dimension $\chi_{b}$, resulting in optimized parameters $\theta^{*}_{b}$. We can then compute the energy given by contracting a PQC with these parameters into an MPS of bond dimension $\chi_{a}$ as 
\begin{equation}\label{eq:Echiab}
    E_{\chi_{a}}(\vec{\theta^{*}_{b}}) = \bra{\psi_{\chi_{a}}(\vec{\theta^{*}_{b}})}H\ket{\psi_{\chi_{a}}(\vec{\theta^{*}_{b}})},
\end{equation}
Note that when $\chi_{a} = 2^{n_{x}n_{y}}$, this quantity yields the exact energy given by the PQC state with the parameters $\theta^{*}_{b}$.

\begin{figure*}[!htbp]
    \centering
    \includegraphics[width=\linewidth]{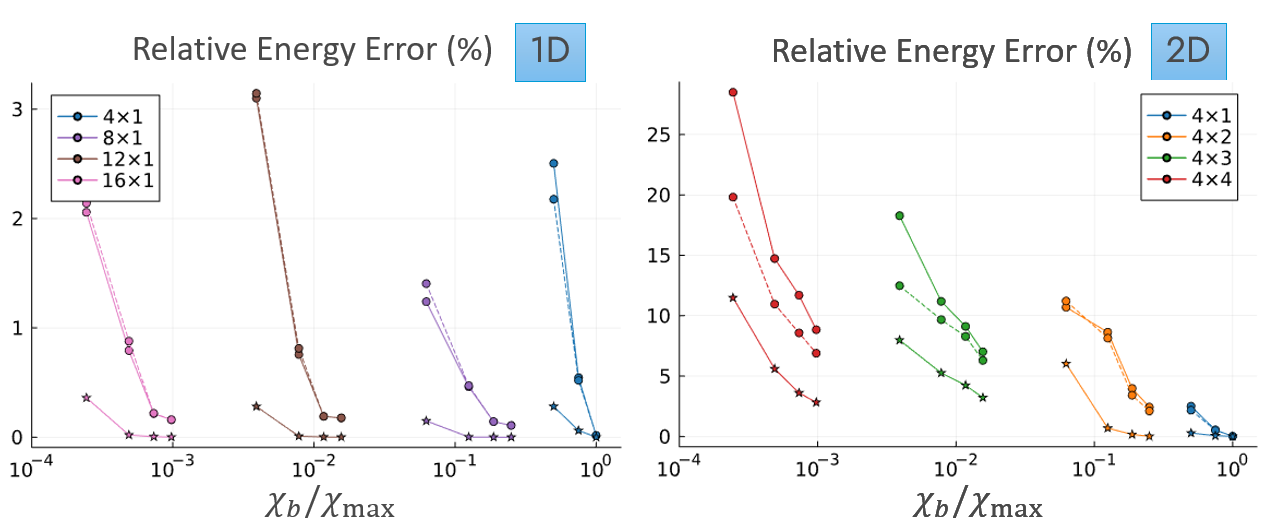}
    \caption{Relative percent energy errors for the Fermi-Hubbard model VQE for different system sizes. The x-axis represents the  bond dimension $\chi_{b}$ that the circuit ansatz is represented as normalized to the maximum bond dimension necessary to exactly represent the circuit for an $n_{x}\times n_{y}$ sized system ($\chi_{\text{exact}} = 2^{n_{x}n_{y}}$). The left plot shows a comparison of 1D Hubbard models, and the right plot shows a comparison of 2D systems. The solid lines represent the VQE relative energy when the circuit is approximated using bond dimension $\chi_{b}$, and the dashed lines represent the VQE relative energy when the circuit is approximated by an MPS of bond dimension $\chi=\text{min}(512, 2^{n_{x}n_{y}})$. DMRG energies (starred points) at these optimization bond dimensions are also shown.}
    \label{fig:energy_errors}
\end{figure*}

Figure~\ref{fig:energy_errors} shows the relative percent energy errors for the Fermi-Hubbard model VQE for different system sizes. Shown on the plots are $E_{\chi_{b}}(\vec{\theta^{*}_{b}})$, $E_{\text{min}(2^{n_{x}n_{y}}, 512)}(\vec{\theta^{*}_{b}})$, and DMRG energies as a function of the optimization bond dimension $\chi_{b}$. For larger system sizes, the exact energies that the optimized energies are being compared to are extrapolated from lower bond dimension DMRG energies.

Our VQE simulations using MPSs display a convergence pattern similar to that of DMRG, albeit with some overhead. We observe that the relative energy errors decrease as the bond dimension increases, highlighting the importance of the bond dimension in obtaining accurate ground state energies. Furthermore, our results indicate that representing the PQC with a larger MPS bond dimension, $\chi_{a} > \chi_{b}$, while retaining the optimization parameters, yields energies that are comparable or even better than those obtained with an MPS bond dimension $\chi_{b}$. In other words, for  $\chi_{a} > \chi_{b}$, we have $ E_{\chi_{a}}(\vec{\theta^{*}_{b}}) \lesssim E_{\chi_{b}}(\vec{\theta^{*}_{b}})$. This feature tells us that when we only have classical access to a bond dimension of $\chi_{b}$, and therefore energy $ E_{\chi_{b}}(\vec{\theta^{*}_{b}})$, the true energy of the ansatz $E_{\chi_{a}}(\vec{\theta^{*}_{b}})$, where $\chi_{a}=2^{n_{x} n_{y}}$ , when ran on quantum hardware, will have equal or lower energies. 

For most 1D Hamiltonians of interest, the convergence of DMRG is very rapid and in practice, large enough bond dimension and system size are accessible numerically~\cite{PhysRevX.5.041041}. In our 1D VQE simulations, we find that, just as in DMRG, we only need a relatively small bond dimension to get close to the ground state. As a result, in figure~\ref{fig:energy_errors} we see a very weak dependence between the relative energy errors and the system size. For all 1D systems (up to 32 qubits), a bond dimension of $\chi_{b} = 16$ gets us a relative energy error of $\lesssim 3\%$.

\begin{figure}[!htbp]
    \centering
    \includegraphics[width=\linewidth]{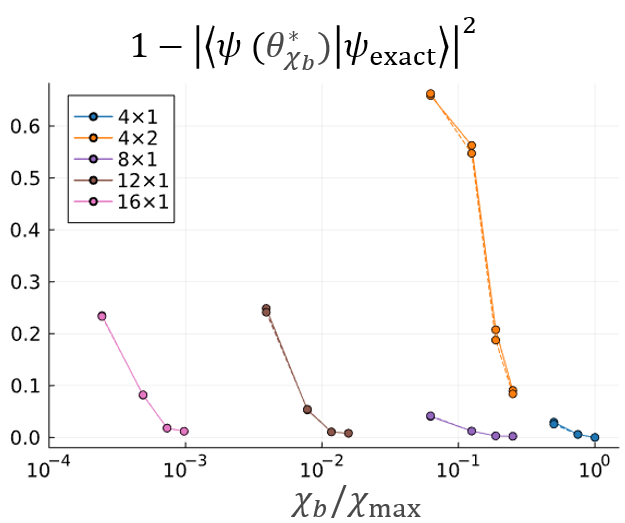}
    \caption{Infidelity measurements between VQE optimized ground states and the exact DMRG obtained ground state.}
    \label{fig:infidelities}
\end{figure}

Focusing on 2D systems, as shown in figure~\ref{fig:energy_errors}, we observe that the relative energy errors tend to increase with larger lattice sizes, indicating that achieving an accurate ground state energy becomes more challenging as the system size grows. This is consistent with our expectations, as larger systems exhibit a more complex entanglement structure, which necessitates a higher bond dimension for accurate representation. Nevertheless, the classical VQE demonstrates a similar trend as DMRG as the bond dimension is tuned. To further assess the performance of our VQE simulations, we show in figure~\ref{fig:infidelities} the fidelity errors for selected system sizes which corroborates the effectiveness of our approach in approximating the ground state of the Fermi-Hubbard model.

\begin{figure*}[!htbp]
    \centering
    \includegraphics[width=\linewidth]{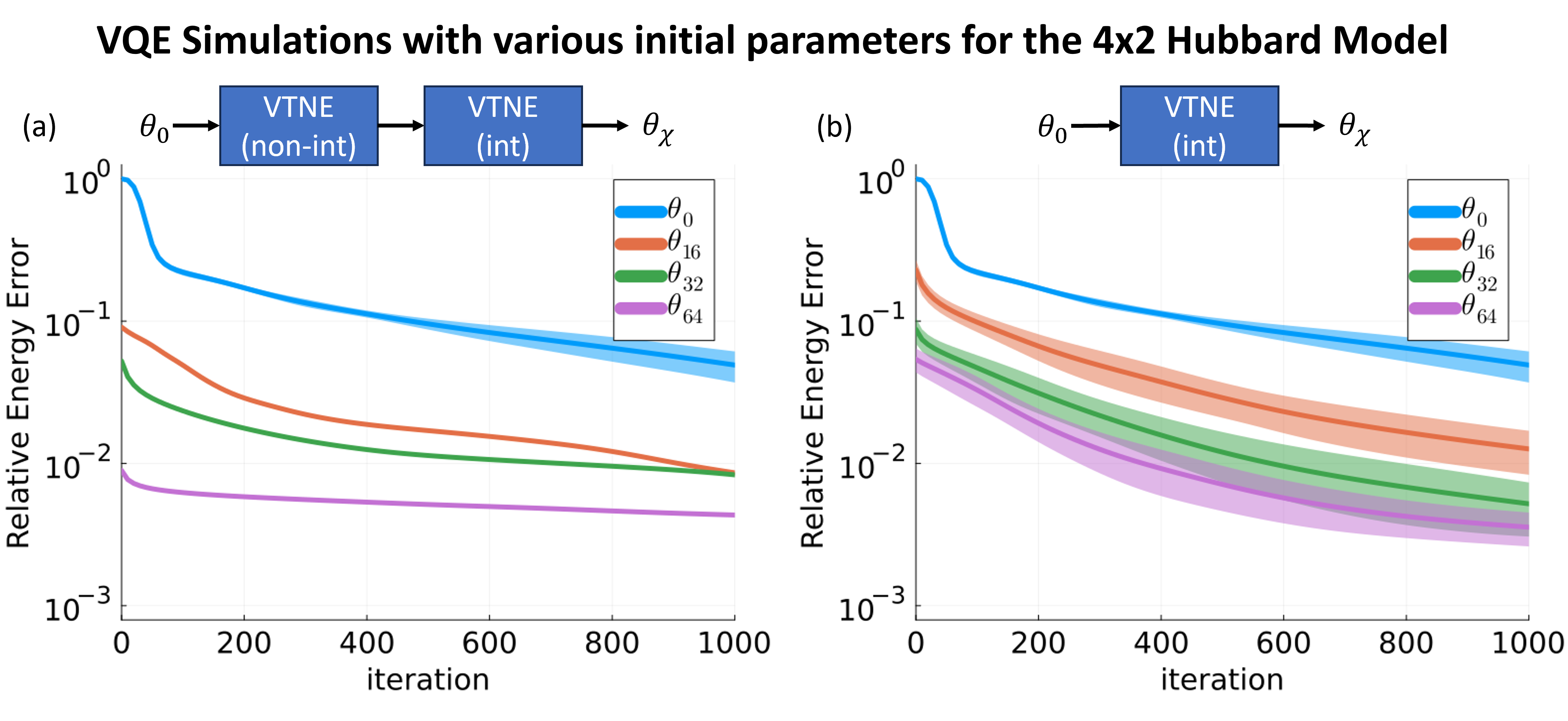}
    \caption{Relative energy error vs optimization step for different parameter initialization methods for the $4\times 2$ Hubbard Model. Shown are the mean relative energy errors at each step, averaged over 10 sampled parameter sets. The blue line represents initial parameters \(\vec{\theta}_{0}\), randomly sampled from a Gaussian distribution \(\mathcal{N}(0, 10^{-3})\). The red, green, and purple lines correspond to optimizations initialized with parameters \(\vec{\theta}_{16}\), \(\vec{\theta}_{32}\), and \(\vec{\theta}_{64}\), respectively. In (a), these parameters are found by first optimizing the non-interacting Hubbard Model first, then the interacting case. In (b), the parameters are obtained by classically optimizing an MPS-assisted VQE simulation at the specified bond dimensions, with the classical optimizer initialized using the parameter set \(\vec{\theta}_{0}\).}
    \label{fig:shots}
\end{figure*}

Most importantly, we study whether doing VQE on top of the parameters that were classically found for the PQC converges faster than starting with random parameters and running VQE without any classical pre-optimization. 
To explore this, we compare parameters found via classical optimization with random initial parameters. Looking at the $4\times 2$ lattice (where $\chi_{\text{max}}=256$), we sample 10 parameter sets $\{\vec{\theta}_{0}^{~(i)}\}_{i=1}^{10}$, with each parameter randomly picked from a Gaussian distribution $\mathcal{N}(0, 10^{-3})$.  
We compare these parameters with the classically optimized parameters $\vec{\theta}_{\chi}$ obtained above. We run a full VQE simulation with the Adam optimizer ($\alpha = 0.001$) using parameter sets $\{\vec{\theta}_{0}^{~(i)}\}_{i=1}^{10}$ and $\vec{\theta}_{\chi_{b}}$. Figure~\ref{fig:shots}a shows the average relative energy error vs the optimization step (or number of gradient calls) for $\vec{\theta}_{0}$, $\vec{\theta}_{16}$, $\vec{\theta}_{32}$, and $\vec{\theta}_{64}$. Overall we find that optimizing classically beforehand with an MPS backend significantly saves the number of gradient evaluations needed to reach lower energies. For example, we save about 400 gradient calls by performing an approximate VQE with a bond dimension 16 MPS and about 1,000 gradient calls using a bond dimension 32 MPS. We also test the robustness of this method by removing the step involving optimizing the non-interacting case instead. That is, we use VTNE to obtain optimized parameters $\vec{\theta}_{\chi_{b}}^{~(i)}$ for each $\vec{\theta}_{0}^{~(i)}$. We find similar advantages shown in figure~\ref{fig:shots}. 
In general, for ansatz where we do not have a good starting point, a large number of computations are needed to match the classically optimized starting configuration. 
This suggests that classical optimization before employing quantum hardware plays an instrumental role in guiding the quantum algorithm toward the global minima more efficiently, which is especially crucial given the current limited quantum resources.  

\section{Discussion}\label{sec:discussion}
It is worth distinguishing our approach from alternative approaches that work directly with the best bond-dimension MPS~\cite{rudolph2022synergy, PhysRevResearch.3.033002, PhysRevA.101.032310}.  In our approach, we find parameters for a given class of parameterized circuits which can be chosen to be shallow or commensurate with the hardware. These parameters can then be utilized to initialize quantum states on the device. Approaches that work directly with the MPS and then add parameterized gates on top of them are inherently forced to work with deeper circuits as  quantum MPS scales quadratically with the bond dimension~\cite{gustafson2023preparing, Huggins_2019, mottonen2005decompositions} and linearly with the system size. To put this in context, an arbitrary MPS of bond dimension $\chi$ consists of $\lceil\log_{2}2\chi\rceil$-qubit unitary operators. Such unitaries require  
\begin{align}\label{eq:cnots}
    \text{CNOTs} &\leq \frac{23}{16}\Delta_{\chi}^{2}\chi^2 - 3\Delta_{\chi}\chi + \frac{4}{3}\\
    &\sim \mathcal{O}(\chi^{2}) \notag
\end{align}

CNOTS, where $\Delta_{\chi} = 2^{\lceil\log_{2}2\chi\rceil - \log_{2}2\chi}$~\cite{1629135}. Put another way, to construct a circuit with a constraint of depth D, then the bond dimension of the MPS must be $\chi \lesssim \mathcal{O}(\sqrt{D/N})$, where $N$ is the system size. Considering Eq.~\ref{eq:cnots}, a $\chi=64$ MPS could require up to 7,660 CNOT gates per site. In addition to this depth dependency, there is also significantly less freedom in choosing the type of circuit architecture being optimized.

With regards to both the approach that maps DMRG MPS states to circuits and  VTNE, increasing bond dimension leads better performance. However, when it comes to implementing the circuits obtained from these algorithms on quantum hardware, other factors also become relevant. VTNE focuses on generic circuits that will often be shallow or tailored to various hardware constraints, while circuits from DMRG-optimized MPSs have different costs with regards to their implementation.  This allows DMRG to achieve states with lower energies for a given bond dimension but at the cost of deeper circuits when the MPS is implemented on quantum hardware. Crucially, if we fix the circuit depth, DMRG-generated MPSs are restricted to bond dimensions that depend on the system size $(\sqrt{D/N})$. This constraint results in higher energy states when compared to VTNE as we increase the system size. Thus, when circuit depth is a limiting factor, VTNE followed by VQE is a competitive and efficient approach for achieving low  energy states.




\section{Conclusions}\label{sec:conclusions}

We have demonstrated that VTNE, which classically pre-optimizes circuits by
approximately simulating VQE using MPS 
significantly aids VQE optimization, with the bond dimension of the MPS tuning how much work (or the number of gradient evaluations) is saved on quantum hardware.  The work here explores ideas that have been discussed in the recent literature~\cite{baek2022say} and provides a complementary approach to other approximate quantum circuit simulations~\cite{mullinax2023largescale,hirsbrunner2023mp2}. For 1D Hamiltonians, the convergence of our method is rapid, with the difference between the energy of the classically optimized parameters and the true ground state energy only depending weakly on system size.  
In this case, most, if not all of the optimization can be performed classically, and in this regime, this method becomes an algorithm for state-preparation on quantum hardware. In contrast, 2D systems exhibit increased relative energy errors with larger lattice sizes due to more complex entanglement structures, emphasizing the necessity of higher bond dimensions for accurate representation. For 2D systems, our algorithm serves as pre-optimization for generating an initial set of parameters for VQE.

Our work illustrates the effectiveness of using an approximate tensor network backend for VQE, facilitating accurate ground state energy estimation and efficient circuit initialization for large system sizes. This approach stands to enhance the scalability and feasibility of VQE on near-term quantum hardware and extends its applicability to a wide range of quantum many-body and chemistry problems.
\section*{Acknowledgements} \label{sec:acknowledgements}
We are grateful for support from NASA Ames Research
Center. We acknowledge funding from the NASA ARMD
Transformational Tools and Technology (TTT) Project.
 A.K. acknowledges support from USRA NASA Academic Mission Services under contract No. NNA16BD14C through participation in the Feynman Quantum Academy internship program.  BKC acknowledges support from the Department of Energy grant DOE DESC0020165.

\bibliographystyle{apsrev4-2}
\bibliography{references}

\end{document}